\documentclass[twocolumn,aps,reprint,amsmath,showpacs,superscriptaddress]{revtex4-1}
\usepackage{graphicx}
\usepackage{dcolumn}   
\usepackage{bm}        
\usepackage{amssymb}   
\usepackage{amsmath}
\usepackage[utf8]{inputenc}
\usepackage{color}
\usepackage{array}
\usepackage{makecell, booktabs, multirow}
\usepackage{stackengine}
\usepackage{hyperref}
\hypersetup{
    colorlinks=true,
    linkcolor=blue,
    filecolor=magenta,      
    urlcolor=cyan,
}

\begin{document}

\title{Nonequilibrium Kondo effect in a graphene-coupled quantum dot in the presence of a magnetic field}

\author{Levente Máthé}
\email[Corresponding author: ]{levente.mathe@itim-cj.ro}
\affiliation{Department of Molecular and Biomolecular Physics, National Institute for Research and Development of Isotopic
and Molecular Technologies, 67-103 Donath, 400293 Cluj-Napoca, Romania}
\affiliation{Faculty of Physics, Babeș-Bolyai University, 1 Kogălniceanu, 400084 Cluj-Napoca, Romania}

\author{Ioan Grosu}
\affiliation{Faculty of Physics, Babeș-Bolyai University, 1 Kogălniceanu, 400084 Cluj-Napoca, Romania}

\begin{abstract}
Quantum dots connected to larger systems containing a continuum of states like charge reservoirs allow the theoretical study of many-body effects such as the Coulomb blockade and the Kondo effect. Here, we analyze the nonequilibrium Kondo effect and transport phenomena in a quantum dot coupled to pure monolayer graphene electrodes under external magnetic fields for finite on-site Coulomb interaction. The system is described by the pseudogap Anderson Hamiltonian. We use the equation of motion technique to determine the retarded Green's function of the quantum dot. An analytical formula for the Kondo temperature is derived for electron and hole doping of the graphene leads. The Kondo temperature vanishes in the vicinity of the particle–hole symmetry point and at the Dirac point. In the case of particle–hole asymmetry, the Kondo temperature has a finite value even at the Dirac point. The influence of the on-site Coulomb interaction and the magnetic field on the transport properties of the system shows a tendency similar to the previous results obtained for quantum dots connected to metallic electrodes. Most remarkably, we find that the Kondo resonance does not show up in the density of states and in the differential conductance for zero chemical potential due to the linear energy dispersion of graphene. An analytical method to calculate self-energies is also developed which can be useful in the study of graphene-based systems. Our graphene-based quantum dot system provides a platform for potential applications of nanoelectronics. Furthermore, we also propose an experimental setup for performing measurements in order to verify our model.
\end{abstract}

\keywords{graphene; Kondo effect; magnetic field; pseudogap Anderson model; quantum dot}

\maketitle

\section{Introduction}
\label{sec:I}
The discovery of graphene \cite{novoselov2004,novoselov2005} has opened new research directions and led to novel graphene-based electronic devices \cite{sun2013,choi2017,zhang2012,han2017}, which employ its unusual physical properties \cite{rozhkov2011,castro2009,kotov2012}. From theoretical considerations, such an electronic device can be considered as a mesoscopic system that can be realized by a molecular junction or a single quantum dot (QD) or many QDs in a particular arrangement coupled to charge reservoirs by metallic \cite{meir1992,meir1991,meir1993,wingreen1994,jauho1994,zimbovskaya2008,roermund2010,zimbovskaya2014,zimbovskaya2015,sahoo2016,hewson2005,chung2008,crisan2015,cronenwett1998,wiel2000,sasaki2009,dong2001,aono1998, hershfield1992,yeyati1993,goldhaber1998,swirkowicz2003}, ferromagnetic \cite{trocha2007,zitko2012,martinek2003,choi2004} or graphene electrodes \cite{aono2013,aono2014,aono2016,isern2019,rodriguez2018}. The choice of electrodes can have a significant effect on the transport properties of the QD system.

In the present work, we study the nonequilibrium Kondo effect and the transport properties in a QD coupled to graphene-based leads by solving the pseudogap Anderson model \cite{aono2013,aono2014,aono2016,zhu2011a,zhu2011b,rodriguez2018,isern2019,glossop2005,wu2016,cheng2013} with the equation of motion (EOM) technique \cite{ lim2013,lacroix1981,kashcheyevs2006}. In our studies, a magnetic field is applied to the QD causing a Kondo resonance splitting, and a finite on-site Coulomb interaction ($U$) is considered resulting in a shift of the main QD energy level. Furthermore, an asymmetric bias voltage is applied to the leads causing the system to be driven out of equilibrium as it lies in the Kondo regime. Our QD system allows the theoretical analysis of quantum many-body effects such as the Coulomb blockade and the Kondo effect, and provides a platform for the implementation of graphene-based nanoelectronic devices.

Graphene is a two-dimensional layer of graphite. Such layers were isolated experimentally about a decade ago \cite{novoselov2004,novoselov2005} and described theoretically about half a century ago \cite{wallace1947}. The carbon atoms in graphene are arranged in a hexagonal (honeycomb) lattice. The hexagonal structure can be considered as a triangular Bravais lattice with a basis of two atoms per unit cell \cite{sarma2011}. Its reciprocal lattice consists of six points at corners of the first Brillouin zone, whereas only two of them are inequivalent. These points are denoted by K and K' and referred to as valleys. At these points the energy dispersion of quasiparticles in graphene is linear in momentum. This linear band structure is called a Dirac cone, and it is at the basis of many interesting physical phenomena such as the `chiral' quantum Hall effect \cite{geim2007}, the Klein tunneling effect \cite{sarma2011} and the Aharonov–Bohm effect \cite{schelter2012}. The points where the conduction and valence bands touch each other in momentum space are the Dirac points. The Dirac points play an important role in the electronic properties of graphene because around these points low-energy excitations can be achieved. Furthermore, the motion of the charge carriers is described by a 2D Dirac-like equation. Therefore, they are often called `relativistic' massless fermions even if they move with a speed which is about 300 times smaller than the speed of light. The density of states (DOS) is proportional to the absolute value of energy \cite{crisan2016}. For all of the above, graphene provides many applications \cite{geim2009} due to its unique mechanical, optical, electronic and thermal properties.

In QD systems, described by an Anderson Hamiltonian, the Kondo effect can be observed in an adequate approximation, e.g., in the Lacroix approximation \cite{lacroix1981}. The Kondo effect is a result of the interaction between a single magnetic impurity (e.g., a magnetic atom or a localized spin) and the free electrons in a nonmagnetic material (e.g., a continuum of states) \cite{cronenwett1998,wiel2000}. The Kondo effect has been extensively studied theoretically within the framework of the EOM method applying the Anderson model to a single QD with metallic contacts \cite{meir1993,zimbovskaya2008,roermund2010,sahoo2016,swirkowicz2003}, which has been confirmed by experimental measurements \cite{cronenwett1998,wiel2000,kogan2004}. This effect is more complicated in a multi-QD system (like a T-shaped double-QD system) with metallic leads due to the presence of multi-electron transport channels \cite{chung2008,crisan2015}, which is also verified by electronic transport measurements \cite{sasaki2009}. The nonequilibrium Kondo effect in QDs affected by finite magnetic fields has been explored in the noncrossing approximation (NCA) \cite{meir1993} and by slave-boson mean-field theory (SBMFT) for finite \cite{dong2001} and infinite Coulomb interaction \cite{aono1998}, respectively. The NCA was also used to study the nonequilibrium Kondo effect in QDs without magnetic fields \cite{wingreen1994}. In addition, the transport properties of QDs were studied within the framework of modified perturbation theory \cite{hewson2005,hershfield1992,yeyati1993}. Moreover, the DOS of a single QD was calculated via EOM and NCA \cite{meir1993}. It was found that in the nonequilibrium DOS the Kondo peaks are located at the values of the chemical potentials and are suppressed due to nonequilibrium dissipation processes. In the presence of a magnetic field, the Zeeman energy shifts the Kondo peaks from the chemical potentials, to the right for spin-up electrons and to the left for spin-down electrons. Thus, the Kondo peaks originally located at the values of the chemical potentials are split into two new peaks. It was found that the differential conductance consists of an observable peak when the asymmetric bias voltage equals the Zeeman energy. In the case of finite Coulomb interaction, the zero-temperature linear conductance is suppressed under a magnetic field, and double-resonant peaks show up at high magnetic fields \cite{dong2001} in good agreement with the experimental measurements \cite{goldhaber1998}. The approximation proposed by Meir et al. \cite{meir1991} was used to analyze the nonequilibrium Kondo effect in QDs for arbitrary Coulomb interaction \cite{swirkowicz2003,zimbovskaya2008,roermund2010} and infinite Coulomb interaction \cite{meir1993,kashcheyevs2006,lim2013}.

The influence of magnetic adatoms on graphene has been studied theoretically \cite{uchoa2008,uchoa2009,uchoa2011,jacob2010,saha2010,wehling2010a,wehling2010b}. It was found that a magnetic moment can be engineered by electrically controlling the properties of a transition metal adatom on graphene providing the possibility to develop graphene-based spintronic devices \cite{uchoa2008}. Therefore, the physical properties of a graphene monolayer with one of its carbon atoms substituted with a magnetic impurity have been explored via density functional calculations \cite{krasheninnikov2009,santos2010}.

In several studies, the Kondo effect in graphene is treated within the framework of magnetic impurities with massless Dirac fermions via a pseudogap Anderson model \cite{anna2010,zhu2010,zhu2011a,zhu2011b,glossop2005,wu2016,cheng2013}. Dell’Anna showed that in graphene, depending on the position of the impurity, within the tight-binding approximation, the number of Kondo couplings can be reduced to one, producing a multichannel pseudogap Kondo model \cite{anna2010}. The Kondo effect was investigated using the tight-binding formalism for different positions of a magnetic adatom on the graphene layer \cite{zhu2010}. The effect of a magnetic adatom located above one carbon atom of graphene was studied by a slave-boson calculation \cite{zhu2011a}. Zhu and Berakdar determined the DOS of a magnetic adatom on graphene in the Kondo regime based on the EOM technique in the Lacroix approximation \cite{zhu2011b}. They found that the Kondo resonance appears only in a narrow energy range for the impurity level with respect to the chemical potential ($\mu$), and the energy scale is proportional to $|\mu|$. The Kondo effect of an adatom on the surface of graphene and its scanning tunneling microscopy (STM) have been analyzed by Keldysh nonequilibrium Green's function theory by Li and co-workers \cite{li2013}. They found that the Kondo peak can be observed in a wide parameter range from the Kondo regime to the mixed valence regime or to the empty orbital regime, which is in agreement with the literature \cite{withoff1990, gonzalez1996,gonzalez1998}. In the latter two regimes, the shape of the Kondo resonance is influenced by the Fano resonance. However, the tunneling between the STM tip and graphene does not obviously affect the shape of the Kondo resonance in the vicinity of zero bias. Yanagisawa investigated the Kondo effect induced by the s–d interaction with Dirac electrons using Green's function theory \cite{yanagisawa2015}. He found that the Kondo temperature is proportional to $|\mu|$ and vanishes at $\mu = 0$.

The numerical renormalization group approach (NRG) was used to analyze the pseudogap Kondo effect of a magnetic impurity in the gapless Kondo \cite{withoff1990,gonzalez1998} and Anderson models \cite{gonzalez1996,gonzalez1998}. It was found that the Kondo resonance is not observed when the Fermi energy is near the Dirac points. The Kondo effect induced by a point defect or a magnetic impurity in graphene has been extensively studied for finite on-site Coulomb interaction within the NRG framework \cite{kanao2012,lo2014}. May et al. studied the Kondo effect for a carbon vacancy in a monolayer of graphene via an effective two-orbital single impurity model using the NRG approach \cite{may2018}. Vojta et al. applied the pseudogap Kondo and Anderson models using a combination of analytical and numerical renormalization group approaches to study the Kondo screening of magnetic impurities in graphene \cite{vojta2010}. They found an asymmetric behavior of the Kondo temperature depending on the sign of the chemical potential in the limit of $U \to \infty$. Experimental measurements of the Kondo effect produced by lattice vacancies in graphene layers reveal high values of the Kondo temperature enabling many future applications \cite{chen2010,ugeda2010}.

A plethora of experimental measurements have focused on the electronic transport through graphene QDs \cite{guttinger2009,guttinger2010,guttinger2011,guttinger2012}, bilayer graphene QDs \cite{eich2018,overweg2018}, graphene QDs in the multilevel regime \cite{muller2012,jacobsen2012,jacobsen2014}, graphene double QDs \cite{molitor2009,molitor2010} and graphene triple QDs \cite{bischoff2013} in the Coulomb blockade regime. The literature concerning transport phenomena in graphene-based QDs described by the pseudogap Anderson model is limited. There are a few theoretical reports based on NCA calculations, which describe the thermoelectric characteristics of a strongly interacting QD connected to electrodes of massless Dirac fermions in the zero-bias voltage limit \cite{aono2013,aono2014}. The zero-bias conductance plots reveal an impurity quantum phase transition between the Kondo and local moment regimes. Furthermore, the thermopower changes its sign when the temperature approaches the Kondo temperature. In a recent study, the thermoelectric properties of a noninteracting QD coupled to massless Dirac fermions have been analyzed using the EOM technique \cite{aono2016}. At low temperature, by tuning the voltage of the metallic gate electrode, this QD system reaches large values of thermopower and figure of merit. Moreover, the thermoelectric properties of a single QD connected to graphene electrodes have been studied within the framework of the Hartree–Fock approximation using the EOM technique, focusing on the Coulomb blockade regime \cite{isern2019}. It was established that the Wiedemann–Franz law is not fulfilled for the graphene contacts. In addition, the thermoelectric transport properties of a noninteracting QD coupled to pure and gapped graphene electrodes have been analyzed based on the Hartree–Fock approximation by the EOM technique \cite{rodriguez2018}. A significant enhancement of the figure of merit was reported for the gapped graphene electrodes within the massless gap scenario. The systems present a high heat-to-electricity conversion efficiency at low temperature, for which the phonon contribution can be neglected \cite{aono2016,isern2019}. 

The analytical approaches to Kondo physics of magnetic impurities in graphene mostly explore the $U \to \infty$ limit. Here, we study the Kondo-type transport properties of a QD connected to graphene leads considering the influence of temperature, magnetic fields and applied bias voltage. An analytical Green's function is derived using the EOM method. We obtain an estimative formula for the Kondo temperature showing similar behavior to previous works. The corresponding DOS and differential conductance are also examined. The paper is organized as follows. First, we introduce the model and determine the Green's function of the QD using the EOM method. Then, we derive formulas for the differential conductance and for the Kondo temperature. Subsequently, we present the results obtained for the DOS and the differential conductance in the absence and the presence of magnetic fields in the Kondo regime. Finally, we summarize the main results of our work.

\section{Model and Analytical Results}
\label{sec:II}
\subsection{Theoretical model}
\label{sec:IIA}
\begin{figure}
\includegraphics[width=8.2cm,keepaspectratio]{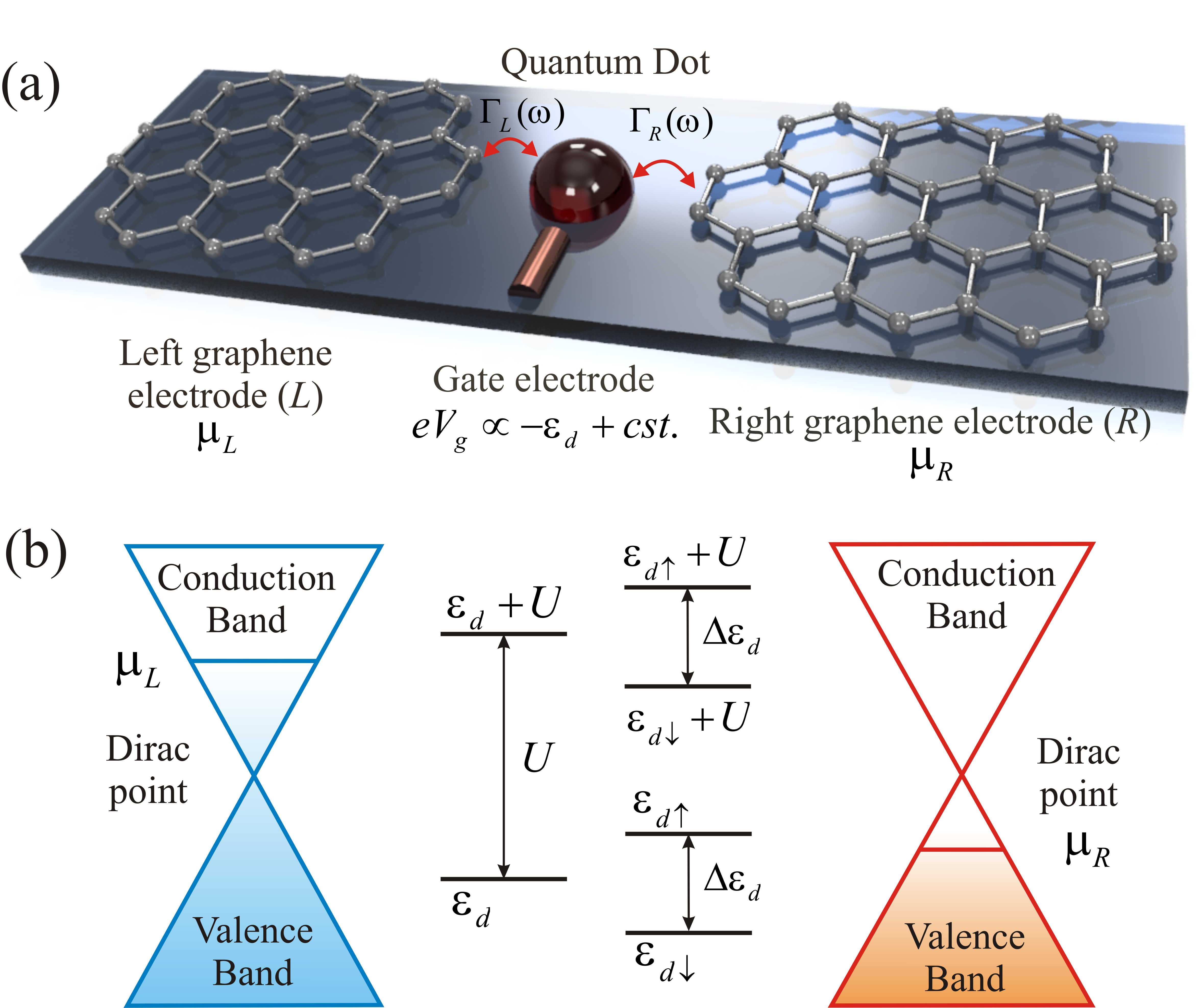}
\caption{(a) Schematic picture of a QD coupled to pure monolayer graphene electrodes at its left (\textit L) and right (\textit R) sides with different chemical potential ($\mu_L$ and $\mu_R$). $\Gamma_{\alpha}(\omega)$ represents the coupling strength between the QD and the $\alpha$ graphene electrode. The spin-independent QD energy level $\varepsilon_d$ can be modulated by the gate voltage $V_g$ at the gate electrode. (b) Schematic representation of the energy level diagram of the QD. The energy levels are shifted by the Coulomb energy ($U$). The applied magnetic field ($B$) leads to a splitting of the discrete energy levels.}
\label{fig:1}
\end{figure}

We consider a QD coupled to pure monolayer graphene electrodes at its left (\textit L) and right (\textit R) sides as shown in Figure \ref{fig:1}. The system can be described by the pseudogap Anderson Hamiltonian \cite{aono2013,aono2014,aono2016,zhu2011a,zhu2011b,rodriguez2018,isern2019,glossop2005,wu2016,cheng2013}:
\begin{equation}
H=H_G+H_D+H_V.
\label{eq:1}
\end{equation}
The first term in Equation \eqref{eq:1}, $H_G$, represents the graphene contact Hamiltonian, which describes the massless Dirac fermions in the left (\textit L) and right (\textit R) graphene leads and can be expressed as:
\begin{equation}
H_G=\sum_{\alpha,s,\sigma}\int_{-k_c}^{+k_c} dk\,\varepsilon_{k}\,c_{\alpha s k\sigma}^{\dagger}c_{\alpha s k\sigma},
\label{eq:2}
\end{equation}
where $c_{\alpha s k\sigma}^{\dagger}$ and $c_{\alpha s k\sigma}$ denote the creation and annihilation operators of the Dirac fermions with wave vector \textit k and spin $\sigma$. Here, $\alpha$ identifies the channel in the left (\textit L) and right (\textit R) graphene leads, and $s$ denotes the valley index. $\varepsilon_k = \hbar v_F k$ is the linear energy dispersion of the Dirac fermions in graphene with $v_F \approx 10^{6}m/s$ being the graphene Fermi velocity. All electrodes are at same temperature $T$. Thus, the momentum distributions of the Dirac fermions inside the graphene electrodes are given by the Fermi–Dirac function
\begin{equation*}
f_\alpha (\varepsilon_{k}) \equiv f(\varepsilon_{k}-\mu_{\alpha})=1/\big[e^{(\varepsilon_{k}-\mu_\alpha)/k_{B}T}+1\big],
\end{equation*}
where $\mu_\alpha$ is the chemical potential of lead $\alpha$, which is assumed to be temperature-independent (in the following $k_B=1$). For every $k$ there is a momentum cutoff ($k_c$) that defines the validity of the linear dispersion and the energy cutoff $D=\hbar v_F k_c$.

The second term in Equation \eqref{eq:1}, $H_D$, models the interacting QD and reads:
\begin{equation}
H_D={\sum_{\sigma} \varepsilon_{d\sigma}d_\sigma^{\dagger}d_\sigma}+{U n_\uparrow n_\downarrow},
\label{eq:3}
\end{equation}
where $\varepsilon_{d\sigma}$ is the discrete spin-dependent energy of the QD, and $d_\sigma^{\dagger}$ and $d_\sigma$ represent the fermionic creation and annihilation operators of the localized electrons in the QD. We assume that an external magnetic field $B$ is applied to the QD resulting in a splitting of the discrete energy level $\varepsilon_{d \sigma}=\varepsilon _d+\sigma \Delta \varepsilon _d/2$ where $\sigma = +1$ and $\bar\sigma = -1$ correspond to spin up and down. The QD energy level $\varepsilon_d$ can be tuned by the gate voltage $V_g$. Therefore, $\Delta \varepsilon_d = |g| \mu_B B$ represents the Zeeman energy of the QD with $\mu _B$ and $g$ being the Bohr magneton and the Landé factor. The second term in Equation \eqref{eq:3} describes the repulsive interaction between the electrons of the QD where $U$ and $n_\sigma=d_\sigma^{\dagger}d_\sigma$ represent the Coulomb repulsion energy and the occupation number operator of the localized electrons with spin $\sigma$. In the absence of a magnetic field, the Coulomb energy $U$ leads to a shift of the main energy level $\varepsilon_d$ resulting in two discrete energy levels $\varepsilon_d$ and $\varepsilon_d+U$. The first level $\varepsilon_d$ corresponds to the singlet state with energy $\varepsilon_d$, and the second is the doublet state with energy $2\varepsilon_d+U$. We consider the situation when $\varepsilon_d$ is well below the values of the chemical potential in the leads ($\mu_\alpha$), while $\varepsilon_d+U$ is situated well above. This setting of energy levels ensures that the graphene-based QD system lies in the Kondo regime.

The last term in Equation \eqref{eq:1}, $H_V$, is the Hamiltonian describing the tunneling between the graphene leads and the QD, which is given by:

\begin{equation}
H_V=\sum_{\alpha,s,\sigma}\int_{-k_c}^{+k_c} dk\big [V_{\alpha s \sigma}(k)c_{\alpha s k\sigma}^{\dagger}d_\sigma+H.c. \big ],
\label{eq:4}
\end{equation}
where $V_{\alpha s \sigma}(k)$ represents the tunneling amplitude, which is related to the coupling strength of the graphene contacts to the QD (see subsection \ref{sec:IIC}). Here, we consider that the graphene monolayer electrodes are arranged in an armchair configuration (see Figure \ref{fig:1}). The distance in real space between the QD and sites of graphene sublattices $A$ and $B$ are approximately the same. In the energy dispersion for armchair edges there is a superposition of the K and K' valleys. Therefore, the tunneling amplitude may be considered valley independent. We also assume that the QD is symmetrically coupled to the leads, while the tunneling amplitude is spin independent. Thus, $ V_{\alpha s \sigma}(k) = V(k)=\tilde V \sqrt{|k|}$, where $\tilde V = V_0 \sqrt{\Omega_0 \pi}/2\pi$ with $V_0$ is the tunneling strength and $\Omega_0$ is the area of the graphene unit cell.

\subsection{Green's function}
\label{sec:IIB}
In order to study the transport properties of the system, the retarded Green's function of the QD needs to be determined. We employ the EOM technique to compute it by applying the broadly used Lacroix decoupling scheme \cite{lacroix1981} within the framework of the Meir approximation \cite{meir1991}. The computational procedure is presented in Supporting Information File 1, Appendix A. We obtain following Green's function of the QD:
\begin{widetext}
\begin{equation}
G_{d\sigma}^{r}(\omega)=\frac{1-\langle n_{\bar{\sigma}}\rangle}{\omega-\varepsilon_{d\sigma}-\Sigma_0^r(\omega)+U\frac{\Sigma_{3\sigma}(\omega)+\Sigma_{4\sigma}(\omega)}{\omega-\varepsilon_{d\sigma}-U-\Sigma_0^r(\omega)-\Sigma_{1\sigma}(\omega)-\Sigma_{2\sigma}(\omega)}}+\frac{\langle n_{\bar{\sigma}}\rangle}{\omega-\varepsilon_{d\sigma}-\Sigma_0^r(\omega)-U-U\frac{\Sigma_{1\sigma}(\omega)+\Sigma_{2\sigma}(\omega)-\Sigma_{3\sigma}(\omega)-\Sigma_{4\sigma}(\omega)}{\omega-\varepsilon_{d\sigma}-\Sigma_0^r(\omega)-\Sigma_{1\sigma}(\omega)-\Sigma_{2\sigma}(\omega)}},
\label{eq:5}
\end{equation}
\end{widetext}
where we define the $\Sigma_0^r(\omega)$ noninteracting tunneling self-energy as:
\begin{equation}
\begin{split}
\Sigma_0^r(\omega)&=\sum_{\alpha ,s}\int_{-k_c}^{+k_c}dk  \frac{V(k)^2}{\omega^+-\varepsilon_{k}}\\
&=-2\eta\bigg(\omega \ln\bigg|\frac{D^2-\omega^2}{\omega^2}\bigg|+i\pi |\omega|\theta(D-|\omega|)\bigg),
\end{split}
\label{eq:6}
\end{equation}
and introduce a dimensionless coupling parameter between the localized electrons and Dirac fermions as $\eta=2{(\tilde V/\hbar v_F)}^2$. The remaining self-energies are given by the relations:

\begin{equation}
\begin{split}
\Sigma_{i\sigma}(\omega)&=\sum_{\alpha ,s}\int_{-k_c}^{+k_c}dk  \frac{V(k)^2}{\Omega_{k\sigma}^{(i)}(\omega)}\\
&=-2\eta\bigg(\omega_{i\sigma} \ln \bigg|\frac{D^2-\omega_{i\sigma}^2}{\omega_{i\sigma}^2}\bigg|+i\pi |\omega_{i\sigma}|\theta(D-|\omega_{i\sigma}|)\bigg),\\
&\quad i=1,2,
\end{split}
\label{eq:7}
\end{equation}
with shorthand notations: $\Omega_{k\sigma}^{(1)}(\omega)\equiv\omega^+-\varepsilon_{k}+(\varepsilon_{d\bar{\sigma}}-\varepsilon_{d\sigma})$, $\Omega_{k\sigma}^{(2)}(\omega)\equiv\omega^++\varepsilon_{k}-(\varepsilon_{d\bar{\sigma}}+\varepsilon_{d\sigma})-U$, $\omega_{1\sigma}=\omega -\sigma \Delta \varepsilon_d$ and $\omega_{2\sigma}=\omega -2\varepsilon_d-U$. Therefore, we have:
\begin{equation}
\Sigma_{3\sigma}(\omega)=\sum_{\alpha ,s}\int_{-k_c}^{+k_c}dk \frac{V(k)^2}{\Omega_{k\sigma}^{(1)}(\omega)}f_\alpha(\varepsilon_{k})=\sum_\alpha\Sigma_{3\sigma}^{\alpha (\gamma)}(\omega),
\label{eq:8}
\end{equation}
\begin{equation}
\Sigma_{4\sigma}(\omega)=\sum_{\alpha ,s}\int_{-k_c}^{+k_c}dk \frac{V(k)^2}{\Omega_{k\sigma}^{(2)}(\omega)}f_\alpha(\varepsilon_{k})=\sum_\alpha\Sigma_{4\sigma}^{\alpha (\gamma)}(\omega).
\label{eq:9}
\end{equation}
Note that an integral similar to the expression defined by Equation \eqref{eq:8} with $\Delta \varepsilon_d=0$ appearing in $\Sigma_{3\sigma}(\omega)$ has been analytically evaluated before \cite{zhu2011b}. We show a simple analytical method for determining the self-energies $\Sigma_{3\sigma}^{\alpha(\gamma)}(\omega)$ and $\Sigma_{4\sigma}^{\alpha(\gamma)}(\omega)$ for finite temperature in Supporting Information File 1, Appendix B. In Appendix C, we compare our analytical results to those of \cite{zhu2011b}. The introduction of the index $\gamma$ in above equations is necessary, because the self-energies $\Sigma_{3\sigma}^{\alpha(\gamma)}(\omega)$ and $\Sigma_{4\sigma}^{\alpha(\gamma)}(\omega)$ have different analytical solutions depending on the sign of the chemical potential. (Here, $\gamma =-, 0$ and $+$ correspond to the cases: $-D<\mu_\alpha\lesssim 0$, $\mu_\alpha=0$ and $0\lesssim\mu_\alpha<D$, see Supporting Information File 1, Appendix A).

The DOS of localized electrons with spin $\sigma$ is given by:
\begin{equation}
\rho_{d\sigma}(\omega)=-\frac{1}{\pi}\textmd{Im} G_{d\sigma}^r (\omega).
\label{eq:10}
\end{equation}

The occupation number of electrons in the QD has to be calculated self-consistently by applying the spectral theorem. At low temperature and equilibrium, the occupation number is given by:
\begin{equation}
\langle n_{\sigma}\rangle=\int_{-D}^{D}d\omega f(\omega)\rho_{d\sigma}(\omega)\approx \int_{-D}^{\mu}d\omega\rho_{d\sigma}(\omega),
\label{eq:11}
\end{equation} 
where $f(\omega)$ is the equilibrium Fermi function with $\mu_\alpha=\mu$. At low temperature and out-of-equilibrium, the occupation number can be expressed as:
\begin{equation}
\begin{split}
\langle n_{\sigma}\rangle &=\frac{1}{2} \int_{-D}^{D} d\omega [f_L(\omega) + f_R(\omega)]\rho_{d\sigma}(\omega)\\
&\approx \frac{1}{2} \bigg[\int_{-D}^{\mu_L}d\omega \rho_{d\sigma}(\omega)+\int_{-D}^{\mu_R}d\omega \rho_{d\sigma}(\omega)\bigg].
\end{split}
\label{eq:12}
\end{equation}

\subsection{Current formulas}
\label{sec:IIC}
In order to avoid all possible consequences caused by the symmetry of the system, we chose an asymmetric bias voltage determined by the chemical potentials as $eV = \mu_L - \mu_R$. The asymmetric bias voltage is also preferred experimentally. The spin-dependent current through the QD is given by \cite{meir1992,jauho1994,meir1991,wingreen1994,meir1993}:
\begin{equation}
I_\sigma = \frac{e}{\hbar}\int_{-D}^{+D}d\omega \frac{\Gamma_L(\omega) \Gamma_R(\omega)}{\Gamma_L(\omega)+\Gamma_R(\omega)} \big[f_L(\omega)-f_R(\omega)\big]\rho_{d\sigma}(\omega),
\label{eq:13}
\end{equation}
where $e$ is the elementary charge and $\Gamma_\alpha(\omega)=2\pi \eta|\omega|\theta(D-|\omega|)$ is the coupling strength between the QD and the graphene electrodes. Writing Equation \eqref{eq:6} in the form $\Sigma_0^r(\omega)=\sum_\alpha [\Lambda_\alpha(\omega)-\frac{i}{2} \Gamma_\alpha(\omega)]$ \cite{jauho1994}, the coupling strength $\Gamma_\alpha(\omega)$ can be straightforwardly determined. The coupling strength can be written in the more conventional form used in the literature, that is $\Gamma_\alpha (\omega) = 2\pi\eta \rho_{\text{gr}}(\omega)$ with $\rho_{\text{gr}}(\omega) = |\omega|\theta(D-|\omega|)$ being the DOS of the graphene leads \cite{li2013}. Therefore, for a system consisting of a magnetic adatom in graphene, $\Gamma_\alpha(\omega)$ does not depend on the bias voltage, which agrees with \cite{zhu2011b,li2013}. The total current is calculated as $I=\sum_\sigma I_\sigma$. The differential conductance is expressed as:
\begin{equation}
\begin{split}
\frac{dI_\sigma}{dV}&=\frac{2e^2}{h}\pi^2 \eta \int_{-D}^{D}d\omega |\omega|\theta(D-|\omega|)\\
&\times\bigg\{\big[f_L(\omega)-f_R(\omega)\big]\frac{\partial\rho_{d\sigma}(\omega)}{\partial (eV)}-\rho_{d\sigma}(\omega)\frac{\partial f_L(\omega)}{\partial\omega}\bigg\}.
\end{split}
\label{eq:14}
\end{equation}
\subsection{Kondo temperature}
\label{sec:IID}
The Kondo temperature $T_K$ defines the energy scale of the system and will be calculated by the method presented in \cite{crisan2015,kashcheyevs2006,crisan2010}. The approximations used in our calculations (see Supporting Information File 1, Appendix A) are quantitatively valid above $T_K$. In the regime $T \le T_K$, the results obtained for relevant quantities such as the Kondo temperature, or the shape of the Kondo resonances are only qualitatively valid \cite{meir1991,martinek2003}. We also note that the EOM method underestimates the Kondo temperature $T_K$ and predicts the absence of the Kondo effect at the particle–hole symmetry point even with metallic leads \cite{roermund2010}. But doing so allows us to investigate the nature of these quantities and to understand the phenomena which take place in such systems \cite{wingreen1994,meir1993}. Thus, below we will estimate an expression for $T_K$ as a function of the sign of $\mu$ for arbitrary values of $U$. First, we investigate the behavior of the Green's function given by Equation \eqref{eq:5} for large values of the Coulomb interaction. The denominator of the second part of the Green's function diverges to infinite when $U \to \infty$ causing the second term to vanish. In this limit, the denominator of the first term of the Green’s function converges to a finite value. Thus the first term reproduces the Green's function in the $U \to \infty$ limit and determines the Kondo temperature. Mathematically, the Kondo temperature is the temperature at which the real part of the denominator of the Green's function of the QD vanishes. The denominator of the first part of the Green's function in equilibrium and in the absence of magnetic fields is denoted by $N(\omega,T)$. The solution of the equation $\textmd{Re}N(\omega=0,T=T_K)=0$ gives the Kondo temperature. In the following, we will work in the low temperature limit and assume that $|\mu|$ is small ($|\mu| \ll D$) in order to eliminate all band effects related to the definition of the energy cutoff \cite{uchoa2008}. Thus, we can write the equation:
\begin{equation}
\begin{split}
\textmd{Re}\Sigma_{3\sigma}^{(\gamma)}&+\textmd{Re}\Sigma_{4\sigma}^{(\gamma)}=-\bigg (  c_1+\frac{c_2^2}{c_1} \bigg )\\
&\times\bigg ( \frac{\textmd{Im}\Sigma_{3\sigma}^{(\gamma)}+\textmd{Im}\Sigma_{4\sigma}^{(\gamma)}}{c_2^{-1} c_1^2+c_2} +\frac{\varepsilon_d+\textmd{Re}\Sigma_{0}^{r}}{U}\bigg ),
\end{split}
\label{eq:15}
\end{equation} 
with
\begin{equation}
\begin{cases}
c_1=\varepsilon_d+U+\textmd{Re}\Sigma_0^r+\textmd{Re}\Sigma_{1\sigma}+\textmd{Re}\Sigma_{2\sigma}\\
c_2=\textmd{Im}\Sigma_0^r+\textmd{Im}\Sigma_{1\sigma}+\textmd{Im}\Sigma_{2\sigma},
\end{cases}
\label{eq:16}
\end{equation}
where $\Sigma_{i\sigma} \equiv \Sigma_{i\sigma}(\omega=0,T=T_K)$ with $i=0,1,2,3,4$. In order to get analytical results, we have to restrict the calculations to the limit of $2T_K<2\varepsilon_d+U-\mu<2D$ that exludes the existence of the particle–hole symmetry point. In the following, by taking into account that the chemical potential in graphene leads $\mu$ can be tuned by various doping techniques, we discuss two different cases concerning $T_K$: (i) $\mu > 0$ that corresponds to electron doping and (ii) $\mu < 0$ that corresponds to hole doping. For $\mu>0$, we obtain from Equation \eqref{eq:15} and Equation \eqref{eq:16}:
\begin{equation}
\begin{split}
T_K^{(+)} &\approx \frac{|2\varepsilon_d+U-\mu|^3}{2e^2} \bigg | \frac{D^2-(2\varepsilon_d+U)^2}{(2\varepsilon_d+U)^2} \bigg | ^{\frac{2\varepsilon_d}{U}}\\
&\times \bigg| \frac{D+2\varepsilon_d+U}{(2\varepsilon_d+U)^2} \bigg |^2 \exp \bigg \{ \frac{1}{\eta}\frac{\varepsilon_d(\varepsilon_d+U)}{U(2\varepsilon_d+U)} \bigg \},
\end{split}
\label{eq:17}
\end{equation} 
and for $\mu<0$ we find:
\begin{equation}
\begin{split}
T_K^{(-)} &\approx \frac{|2\varepsilon_d+U-\mu|^3}{2e^{2} |D+2\varepsilon_d+U|^2} \bigg | \frac{D^2-(2\varepsilon_d+U)^2}{(2\varepsilon_d+U)^2} \bigg | ^{-\frac{2\varepsilon_d}{U}}\\
&\times \exp \bigg \{ -\frac{1}{\eta}\frac{\varepsilon_d(\varepsilon_d+U)}{U(2\varepsilon_d+U)} \bigg \},
\end{split}
\label{eq:18}
\end{equation}
where we used the properties of the Lambert W function, and $e$ denotes now the Euler's constant.

To the best of our knowledge, so far, no analytical formula has been derived for the Kondo temperature of the system considered here. In order to verify the correctness of Equation \eqref{eq:17} and Equation \eqref{eq:18}, we compare them to the relation used in another study of a QD with metallic contacts, which is based on the EOM method \cite{roermund2010}:
\begin{equation}
T_K = (2\varepsilon_0+U)\exp \bigg\{ \frac{2\pi \varepsilon_0 (\varepsilon_0+U)}{\Gamma U} \bigg\},
\label{eq:19}
\end{equation}
where $\varepsilon_0=\varepsilon_d-\mu$. We observe that $T_K$ only depends on $U$ and is independent of $D$ for metallic contacts. For graphene contacts, $T_K^{(\gamma)}$ is regulated by $U$ and also by $D$. The presence of $D$ in $T_K^{(\gamma)}$ is due to the fact that it determines the band structure of graphene. Therefore, at the particle–hole symmetry point ($2\varepsilon_d+U=0$), for $\mu \neq 0$, $T_K$ does not vanish for metallic leads within the Lacroix approximation. In our case, we do not exactly reach the particle–hole symmetry point due to the conditions outlined above, but we can estimate $T_K^{(\gamma)}$ in the vicinity of particle–hole symmetry. If we consider that $2\varepsilon_d+U \equiv \alpha$ where $|\alpha| \ll D$, then for $\mu \equiv \mu^{(+)} > 0$ we have:
\begin{equation}
T_K^{(+)} \approx \frac{|\alpha - \mu^{(+)}|^3}{2e^2 \alpha^2} \exp \bigg \{\frac{\alpha^2 - U^2}{4 \eta \alpha U} \bigg\}
\label{eq:20}
\end{equation}
and for $\mu \equiv \mu^{(-)} < 0$ we find:
\begin{equation}
T_K^{(-)} \approx \frac{|\alpha - \mu^{(-)}|^3}{2e^2 \alpha^2} \exp \bigg \{-\frac{\alpha^2 - U^2}{4 \eta \alpha U} \bigg\}.
\label{eq:21}
\end{equation}
We can see that $T_K^{(\gamma)}$ does not vanish in vicinity of the particle–hole symmetric point, in agreement with what was obtained for metallic electrodes. It follows that $T_K$ vanishes only for $\mu =0$ in the case of metallic leads. In our case, if we assume that $\mu^{(+)} \to 0^+$ and $\alpha \to 0^+$ as well as $U>|2\varepsilon_d|$, then from Equation \eqref{eq:20} we have $T_K^{(+)} \approx 0$. In the same way, assuming that $\mu^{(-)} \to 0^-$ and $\alpha \to 0^-$ with $U<|2\varepsilon_d|$, then from Equation \eqref{eq:21} we derive that $T_K^{(-)} \approx 0$. It was shown that the Kondo temperature vanishes at the Dirac points \cite{withoff1990,vojta2010,aono2013,yanagisawa2015}. The NRG results of a pseudogap Anderson model show no Kondo screening in the case of particle–hole symmetry when $\mu = 0$ due to the vanishing DOS at the Dirac point. In the case of particle–hole asymmetry, (with $U \to \infty$) Kondo screening occurs when the QD–lead coupling constant is larger than a critical value \cite {gonzalez1998}. Furthermore, employing NRG calculations Kanao et al. established that the Kondo temperature does not vanish at $\mu = 0$ due to the particle–hole asymmetry, while it is symmetric in $\mu$ in the vicinity of the Dirac point \cite{kanao2012}. In another work, NRG calculations predict a non-zero but asymmetrical behavior for $T_K$ in particle–hole asymmetry \cite{lo2014}. In the $U \to \infty$ limit, by using a combination of analytical and numerical renormalization-group techniques, Vojta et al. found two completely different behaviors for $T_K$ as a function of the sign of $\mu$ \cite{vojta2010}. For $\mu>0$, the Kondo temperature fulfills the law $T_K \propto |\mu|^x$ with a universal exponent $x \approx 2.6$. For $\mu<0$, they found that $T_K \propto \kappa |\mu|$ where $\kappa$ is a universal prefactor. An extreme asymmetry between cases is observed when the graphene is differently doped. Moreover, Yanagisawa deduced a formula for the Kondo temperature using the Green's function theory \cite{yanagisawa2015}. Here, for small values of $|\mu|$, the Kondo temperature is independent of the sign of the chemical potential, namely $T_K \propto |\mu|$. In the $U \to \infty$ limit, an asymmetrical behavior of $T_K$ was reported in case the chemical potential is situated below or above the Dirac points of graphene with a magnetic adatom \cite{li2013}.

In conclusion, the vanishing of $T_K^{(\gamma)}$ at the Dirac point in the vicinity of the particle–hole symmetry point is most likely a consequence of the EOM method. On the other hand, in the case of particle–hole asymmetry, the Kondo temperature $T_K$ does not vanish even at $\mu=0$, which is in accordance with NRG calculations for finite $U$ \cite{kanao2012,lo2014}. The correct dependence of $T_K$ on $\mu$ is still not clear even for $U \to \infty$. Future studies will be necessary to further understand the behavior of $T_K$, also for finite $U$. The precise determination of $T_K$ demands self-consistent calculations instead of working within the Meir approximation to derive average values of the mixing operators $\langle d_{\bar{\sigma}}^{\dagger}c_{\alpha sk \bar{\sigma}}\rangle$ and $\langle c_{\alpha's'k'\bar{\sigma}}^{\dagger}c_{\alpha sk \bar{\sigma}}\rangle$ which appear in the retarded Green's function (see Supporting Information File 1, Appendix A). The Green's function has to be calculated taking into account new higher-order correlation functions resulting from a more accurate decoupling scheme \cite{roermund2010}. This decoupling scheme yields a $T_K$ that does not vanish at the particle–hole symmetry point within the EOM method of QDs with metallic electrodes.
\section{Results and Discussion}
\label{sec:III}
In this section, we present our numerical results on the graphene-based QD system. In order to simplify the numerical calculations, we measure all quantities in units of the energy cutoff $D$. We assume $D \approx 7 \text {\,eV}$, which is an acceptable value for graphene samples \cite{rodriguez2018,uchoa2008,uchoa2009,uchoa2011}. We consider a lateral QD formed by lithographically-realized metallic gate electrodes placed near the QD, similar to a QD created in a semiconductor heterostructure consisting of a two dimensional electron gas. The energy of the confined electrons can be changed using a gate electrode \cite{goldhaber1998}. The Coulomb repulsion energy in the QD is assumed to be $U \approx 480 \text {\,meV}$, i.e., $U/D = 0.069$, an acceptable value for magnetic impurities in graphene \cite{uchoa2008,uchoa2009,uchoa2011}. To estimate the value of the dimensionless coupling parameter $\eta$, we assume a tunneling strength of $V_0 \approx 1 \text{\,eV}$ \cite{uchoa2011,zhu2011b}, and consider that the area of the graphene unit cell is $\Omega_0 \approx 0.051 \text{\,nm}^2$ \cite{sarma2011}. We thus find $\eta \approx 0.0186$, and in the following, we use the rounded value $0.0186$. Importantly, the coupling strength $\eta$ is the main parameter of the graphene-based QD system determining the transport regime. Namely, when $\eta$ gradually decreases, a transition from the Kondo regime to the Coulomb blockade regime of the transport through the QD occurs \cite{mathe2019}. This tendency has been previously explored for a QD attached to metallic electrodes \cite{zimbovskaya2008,zimbovskaya2014,zimbovskaya2015}. First, we present our results for infinite $U$ because they allows us to understand the phenomena which take place in the system. Then, we generalize these results for finite $U$.

\begin{figure}
\includegraphics[width=8.2cm,keepaspectratio]{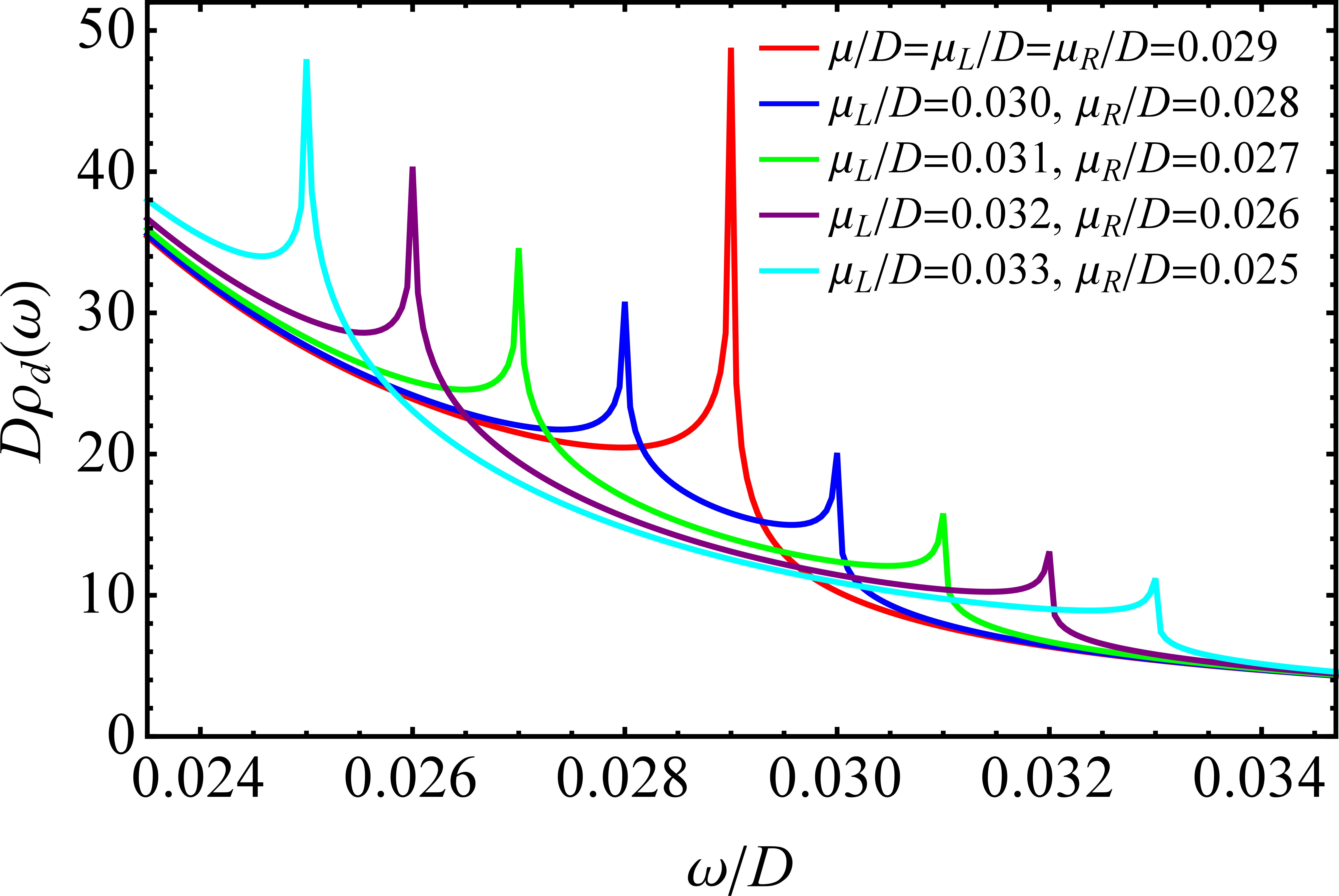}
\caption{The DOS in the QD for different values of the chemical potentials at temperature $T/D = 5\times 10^{-6}$ with $U/D \to \infty$ and in absence of a magnetic field. Additional parameters: $\eta = 0.02$ and $\varepsilon_d /D= -0.011$.}
\label{fig:2}
\end{figure}

In Figure \ref{fig:2}, we present the self-consistent calculations of the total DOS of the QD [sum of the spin-dependent densities of states $\rho_d(\omega) = \rho_{d\uparrow}(\omega) + \rho_{d\downarrow}(\omega)$] for different values of the chemical potentials and at zero magnetic field. At equilibrium ($\mu_L-\mu_R=eV=0$), we observe the Kondo peak in the DOS at the equilibrium chemical potential $\mu_L=\mu_R=\mu$ that corresponds to a resonant transmission through the QD \cite{meir1993}. Applying a bias voltage between the left and right electrodes (at nonequilibrium, $eV\not=0$) results in the splitting of the Kondo peak into two peaks of smaller amplitudes, which are located at the chemical potential of the left ($\mu_L$) and the right ($\mu_R$) electrodes. Thus, the distance between the nonequilibrium Kondo resonances is equal to the difference of the values of the chemical potentials, $\Delta \mu=eV$. The suppression of the peak amplitudes is caused by the dissipative transitions of electrons between the electrode of high chemical potential and the electrode of low chemical potential. This electron transfer determines a finite relaxation time $\tau_{\sigma}$, that can be calculated using second-order perturbation theory \cite{wingreen1994,meir1993}. It can be intuitively introduced by substituting $\delta$ in the self-energies $\Sigma_{i\sigma}(\omega)$ with $\hbar/\tau_{\bar \sigma}$ \cite{swirkowicz2003,martinek2003}. By considering a small value for $\tau_{\bar \sigma}$ e.g., $\hbar/\tau_{\bar\sigma} = \delta \approx 10^{-7}$ \cite{swirkowicz2003}, the analytical results for the self-energies $\Sigma_{i\sigma}(\omega)$ are in good agreement with the numerical calculations (see Supporting Information File 1, Appendix C for a particular case).

\begin{figure*}
\centering
\includegraphics[width=\textwidth]{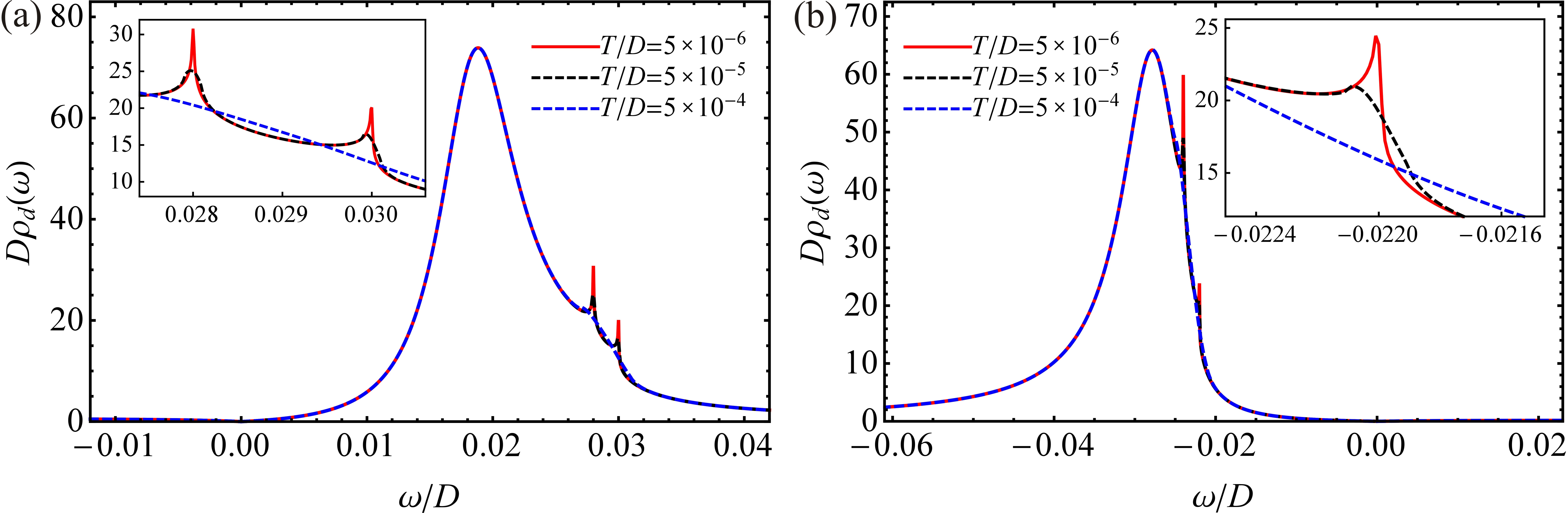}
\caption{The nonequilibrium DOS in the QD at zero magnetic field for different temperatures with $U/D \to \infty$ and different values of the parameters: (a) $\mu_L/D = 0.028$, $\mu_R/D = 0.030$, $\eta = 0.02$ and $\varepsilon_d/D = -0.011$. (b) $\mu_L/D = -0.024$, $\mu_R/D = -0.022$, $\eta = 0.015$ and $\varepsilon_d/D = -0.068$.}
\label{fig:3}
\end{figure*}

The numerical results for the nonequilibrium DOS in the QD calculated for zero magnetic field, three different temperatures and different values of the additional parameters are shown in Figure \ref{fig:3}. At low temperature, the Kondo resonances appear at the values of the chemical potentials ($\omega \approx \mu_L$ and $\omega \approx \mu_R$), and their amplitudes decrease with increasing temperature, completely vanishing at high temperature. These findings are in agreement with what was observed for a QD with metallic electrodes \cite{wingreen1994,meir1993}. At temperatures close to the Kondo temperature or lower, the shapes of the Kondo peaks are not quantitatively valid due to the approximations used here. In order to quantitatively describe the shape of the resonances one should use an approximation based on self-consistent calculations \cite{lacroix1981}. Besides the strongly temperature-dependent sharp Kondo peaks, the DOS consists of a broadened peak with large amplitude, the peak of which is located at $\omega \approx \varepsilon_d + \textmd{Re}\Sigma_0^r(\omega)+\textmd{Re}\Sigma_{3}(\omega)$ where $\textmd{Re}\Sigma_{3}(\omega)=\textmd{Re}\Sigma_{3\uparrow}(\omega)=\textmd{Re}\Sigma_{3\downarrow}(\omega)$. This broadened peak corresponds to a resonant transmission through the QD at $\omega \approx \tilde{\varepsilon}_d$, for which the renormalized QD energy level $\tilde{\varepsilon}_d$ is calculated self-consistently from the relation $\tilde{\varepsilon}_d=\varepsilon_d + \textmd{Re}\Sigma_0^r(\tilde{\varepsilon}_d)+\textmd{Re}\Sigma_{3}(\tilde{\varepsilon}_d)$ \cite{martinek2003}.

\begin{figure*}
\includegraphics[width=\textwidth]{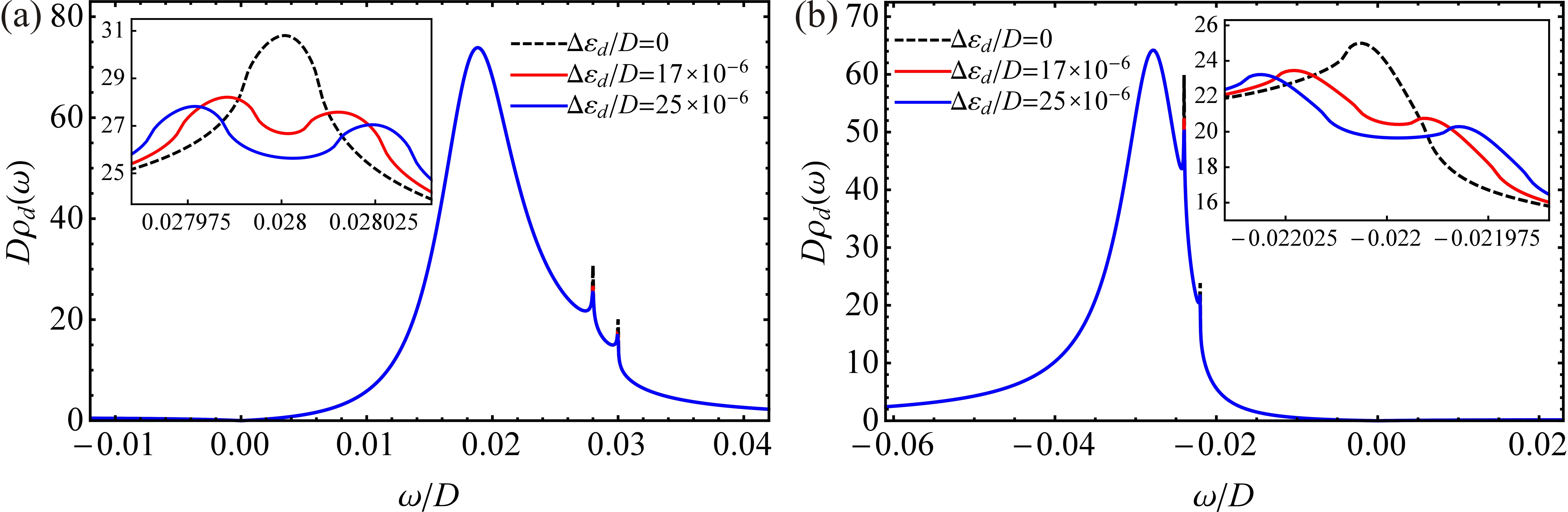}
\centering
\caption{The nonequilibrium DOS in the QD for different values of the Zeeman energy with $U/D \to \infty$ at temperature $T/D=5\times 10^{-6}$ and for different parameters: (a) $\mu_L/D = 0.028$, $\mu_R/D = 0.030$, $\eta = 0.02$ and $\varepsilon_d/D = -0.011$. (b) $\mu_L/D = -0.024$, $\mu_R/D = -0.022$, $\eta = 0.015$ and $\varepsilon_d/D = -0.068$.}
\label{fig:4}
\end{figure*}

We now investigate the effect of an external magnetic field acting only on the QD without affecting the graphene leads. Neglecting the effect of the magnetic field on the electrodes is a broadly used approximation for QDs connected to metallic leads \cite{meir1993,dong2001,utsumi2005}. We can apply this approximation because the graphene leads are well separated. In addition, we consider a magnetic field that is well focused on the QD. Taking into account the modification of the current through the QD as a result of the magnetic field applied to the leads, is beyond the scope of this work. We estimate the Zeeman splitting according to the experimental data. The Landé factor of the QD is assumed $g=-0.44$. For QDs formed in GaAs, this value is close to the $g$ factor of a two-dimensional electron gas \cite{hanson2003,kogan2004,allison2014}. The Bohr magneton is $\mu_B = 58 \, \mu\text{eV/T}$ \cite{kogan2004}. For the applied magnetic field we choose values close to $B \approx 5\text{\,T}$, which is preferable for experimental measurements \cite{hanson2003,kogan2004,allison2014}. Thus, the Zeeman splitting becomes $\Delta \varepsilon_d \approx 125 \,\mu\text{eV}$, which corresponds to $\Delta \varepsilon_d /D \approx 17 \times 10^{-6}$. The values of the chemical potentials remain unchanged. By applying a magnetic field (Figure \ref{fig:4} a,b, out-of-equilibrium), the Kondo peak splits into two peaks of smaller amplitudes, which are shifted by the Zeeman energy from the chemical potential, to the right for spin-up electrons and to the left for spin-down electrons. These results are in agreement with the results of Meir et al. concerning a QD with metallic electrodes influenced by a magnetic field \cite{meir1993}. The location of the damped peak of electrons with spin $\sigma$ is given by $\omega \approx \varepsilon_{d\sigma}+\textmd{Re}\Sigma_0^r(\omega)+\textmd{Re}\Sigma_{3\sigma}(\omega)$. It is shifted from the nonmagnetic position (see Figure \ref{fig:3}) to the right by $\Delta \varepsilon_d/2$ for spin-up electrons and to left for spin-down electrons. The broadened peak corresponds to resonant tunneling of electrons with spin $\sigma$ at the spin-dependent renormalized QD energy level $\tilde{\varepsilon}_{d\sigma}=\varepsilon_{d\sigma} + \textmd{Re}\Sigma_0^r(\tilde{\varepsilon}_{d\sigma})+\textmd{Re}\Sigma_{3\sigma}(\tilde{\varepsilon}_{d\sigma})$. We observe that, in all cases, the DOS totally disappears at $\omega = 0$. In addition, for $\mu_\alpha = 0$ the Kondo peak does not show up in the DOS. This behavior of the DOS is agreement with the results of Li et al. \cite{li2013} for an adatom on the surface of graphene with $U \to \infty$, and is a consequence of the zero DOS of graphene at the Dirac points.

\begin{figure}
\includegraphics[width=8.2cm,keepaspectratio]{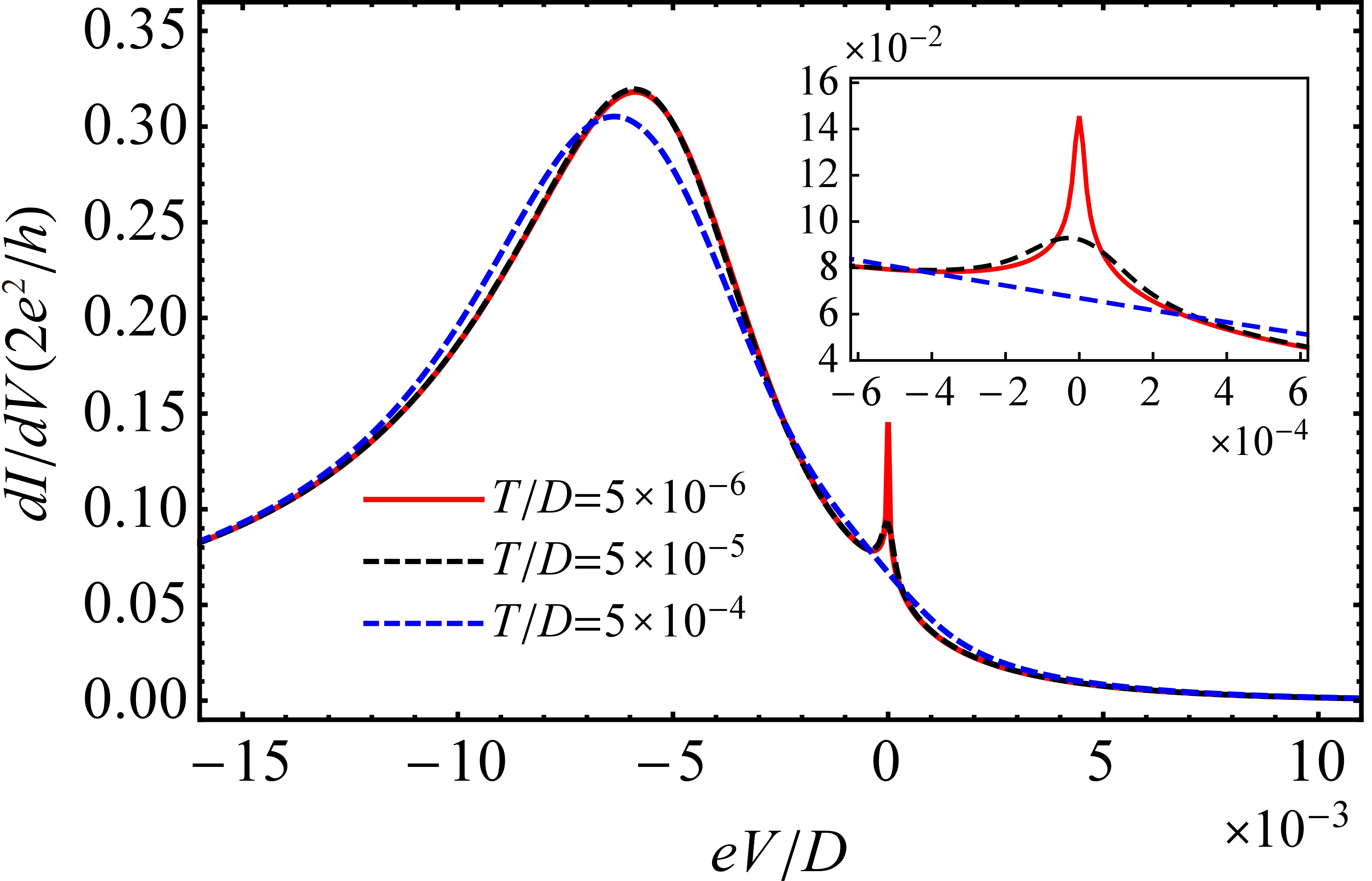}
\caption{Differential conductance $dI/dV$ as a function of the bias voltage $eV$ with $\mu_R/D=-0.022$ at three different temperatures in absence of a magnetic field. The remaining parameters are the same as those in Figure \ref{fig:3}b.}
\label{fig:5}
\end{figure}

In Figure \ref{fig:5}, we present the results obtained for the total differential conductance of the QD [sum of the spin-dependent differential conductance $dI/dV= \sum_\sigma dI_\sigma/dV$] as a function of the bias voltage ($eV=\mu_L-\mu_R$) at three different temperatures and zero magnetic field. The zero-bias peak is observed when the difference in chemical potentials is equal to the Zeeman energy, $eV=\Delta \mu = \Delta \varepsilon_d=0$. The zero-bias peak in the differential conductance corresponds to a resonant transmission through the QD and is strongly temperature-dependent. At low temperature, its shape is narrow and sharp, but with increasing temperature the amplitude decreases, and the peak becomes broadened. We also observe that the shape of the Kondo peak strongly depends on the bias voltage. Its amplitude reaches a maximum at $eV=0$ and quickly drops when $eV$ increases. The variation of the differential conductance near $eV=0$ can be described by a second-degree polynomial function. The dependence of the differential conductance on the bias voltage is in agreement with the theoretical results of Świrkowicz et al., who considered a QD connected to metallic electrodes \cite{swirkowicz2003}. The zero-bias peak of the differential conductance was observed experimentally for magnetic impurities induced by vacancies on graphite surface \cite{ugeda2010}.

\begin{figure}
\includegraphics[width=8.2cm,keepaspectratio]{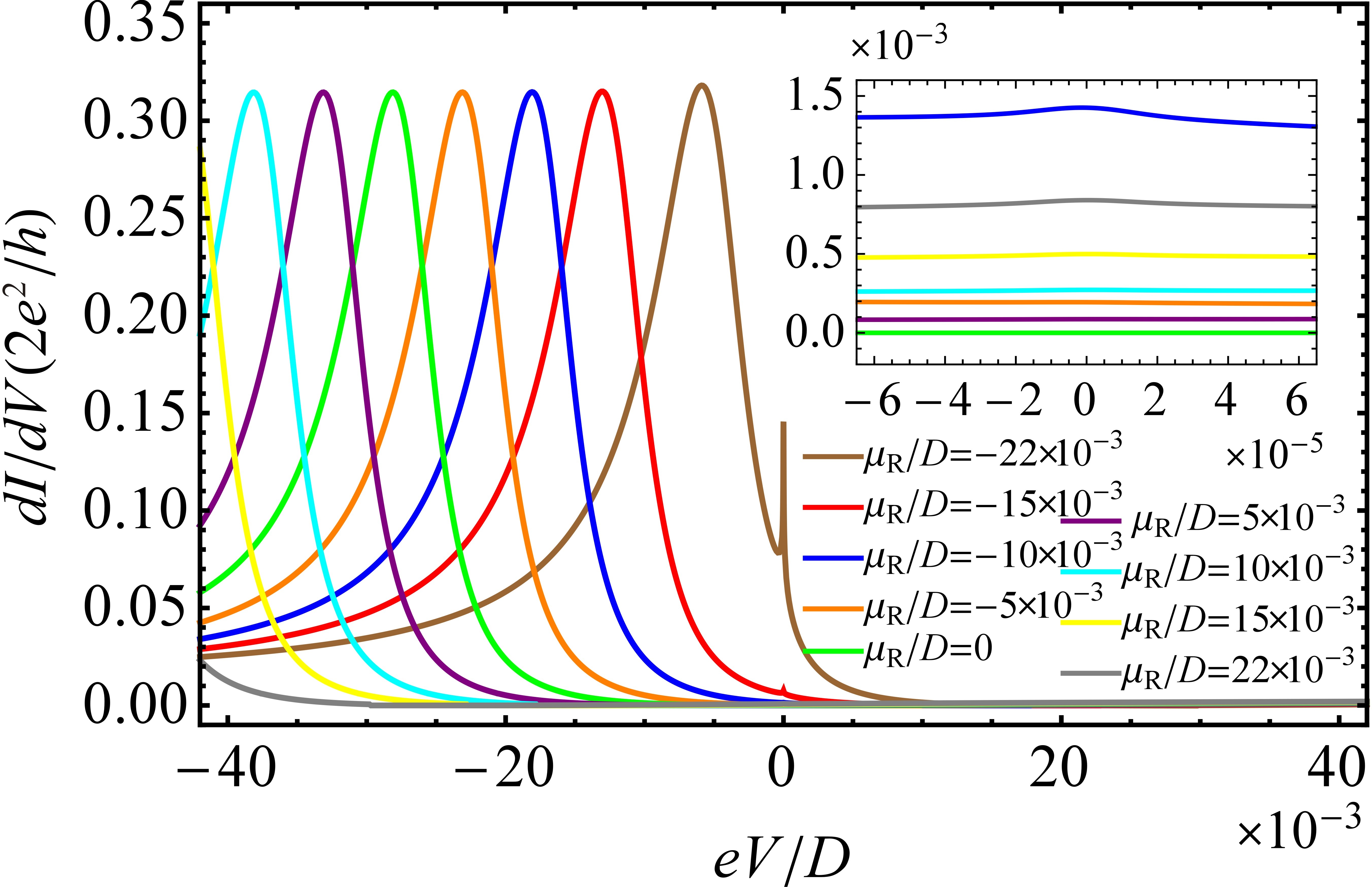}
\caption{Differential conductance $dI/dV$ as a function of the bias voltage $eV$ for different values of the chemical potential $\mu_R$ at temperature $T/D=5\times 10^{-6}$ in the absence of a magnetic field. The remaining parameters are the same as those in Figure \ref{fig:5}.}
\label{fig:6}
\end{figure}

In Figure \ref{fig:6}, we plot the differential conductance as a function of the bias voltage for different values of the chemical potential of the right lead, $\mu_R$, at temperature $T/D = 5 \times 10^{-6}$ and in absence of a magnetic field. Once $\mu_R$ meets the Dirac points, the zero-bias peak vanishes since the DOS is zero at the Dirac points. When $\mu_R$ is situated below or above the Dirac points, the zero-bias peak appears in the differential conductance, revealing the existence of the Kondo effect. When $\mu_R$ is situated below the Dirac points, the zero-bias peak is higher for larger values of $|\mu_R|$. In case $\mu_R$ is above the Dirac points, the zero-bias peak is much smaller than for negative values of $\mu_R$. The difference in height of the zero-bias peaks as a function of $\mu_R$ results from the particle–hole asymmetry of the graphene electrodes. The behavior of the Kondo peak is in agreement with other theoretical results obtained for magnetic impurities in graphene \cite{li2013,vojta2010}.

\begin{figure}
\includegraphics[width=8.2cm,keepaspectratio]{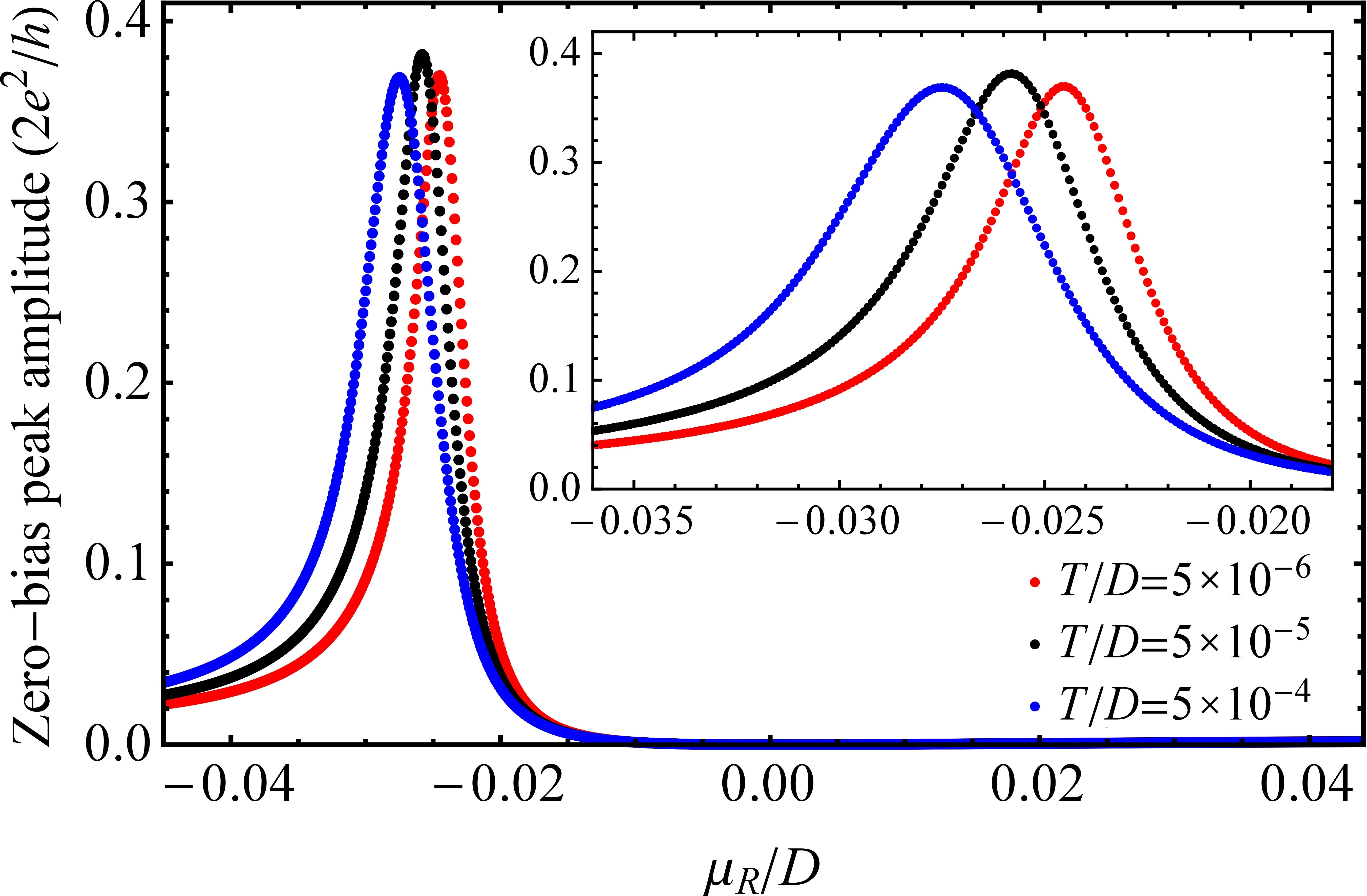}
\caption{Amplitude of the zero-bias peak as a function of the chemical potential $\mu_R$ for $eV=0$ at three different temperatures and in the absence of a magnetic field. The remaining parameters are the same as those in Figure \ref{fig:5}.}
\label{fig:7}
\end{figure}

Figure \ref{fig:7} shows the zero-bias peak as a function of the chemical potential $\mu_R$ for $eV = 0$ at three different temperature values. The figure can be interpreted in the following way. We vertically intersect the graphic at a given point of $\mu_R$, e.g., at $\mu_R/D=-0.022$, and follow the evaluation of the amplitude of the zero-bias peak. Obviously, from the top of the graphic to the bottom, we cross the first curve corresponding to the temperature $T/D=5\times 10^{-6}$, then the second one that corresponds to $T/D = 5\times 10^{-5}$, and finally reach the third curve corresponding to $T/D=5\times 10^{-4}$. The value of the zero-bias peak amplitude at the crossing points decreases with increasing temperature for a narrow range of the chemical potential, when the system is in the Kondo regime, in agreement with the results of Figure \ref{fig:5}. The amplitude of the zero-bias peak strongly depends on $\mu_R$. At $\mu_R=0$ the zero-bias peak totally vanishes and reaches a maximal value for $\mu_R/D = \mu_R^{\text{max}}/D\approx -0.0245$, at $T/D = 5\times 10^{-6}$. With increasing temperature the peak shifts to the left, i.e., to the smaller values of $\mu_R$. For positive values of $\mu_R$, the zero-bias peak has a much smaller amplitude than for negative values of $\mu_R$ (see Figure \ref{fig:6}). Consequently, in order to reach the highest transmission probability of the charge carriers at $eV=0$ with the chosen parameters of the system, we have to set $\mu_R=\mu_R^{\text{max}}$.

\begin{figure}
\includegraphics[width=8.2cm,keepaspectratio]{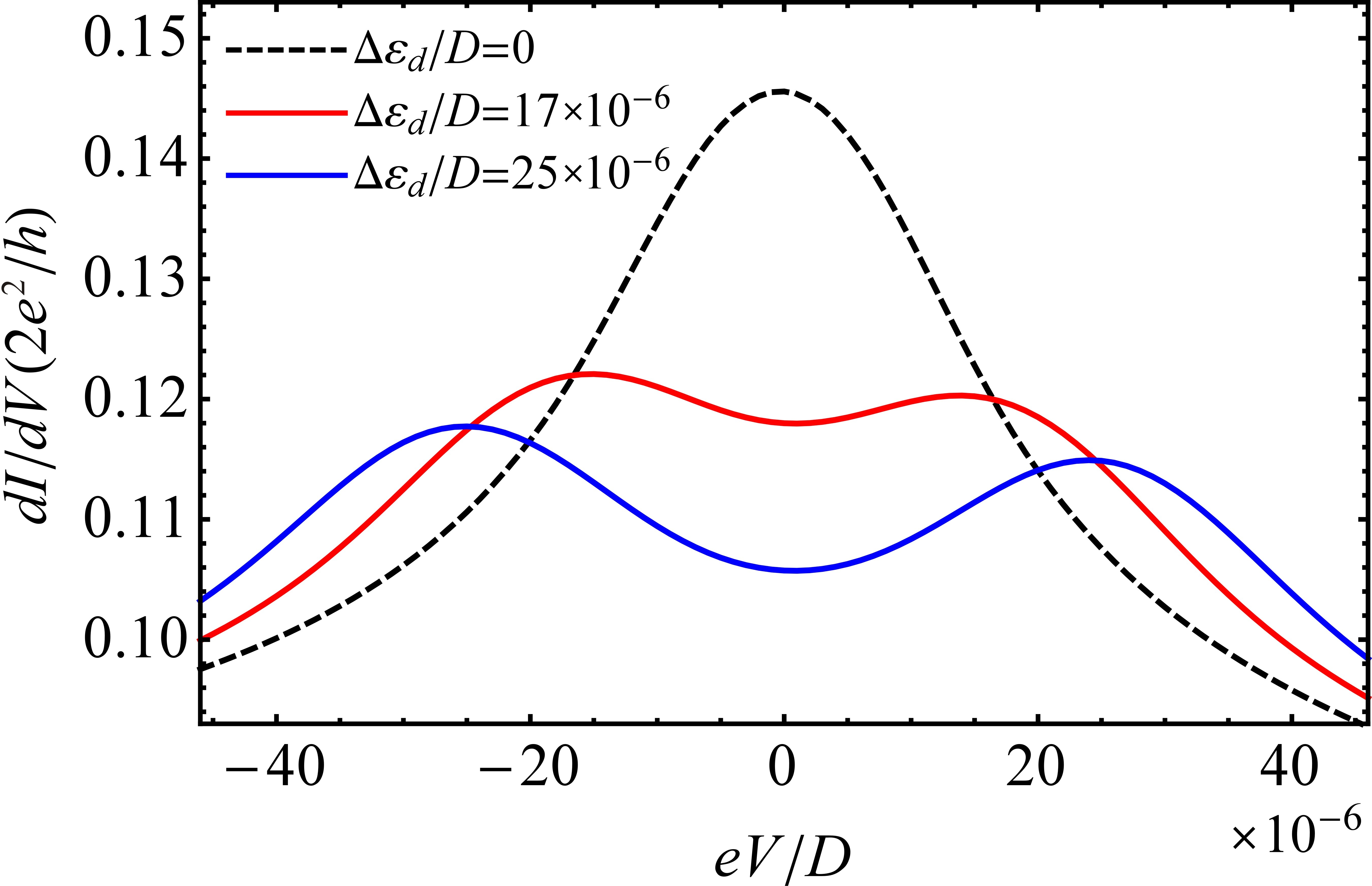}
\caption{Differential conductance $dI/dV$ as a function of the bias voltage $eV$ for different values of the Zeeman energy at temperature $T/D=5\times 10^{-6}$. The remaining parameters are the same as those in Figure \ref{fig:5}.}
\label{fig:8}
\end{figure}

In Figure \ref{fig:8}, we plot the differential conductance as a function of the bias voltage for different values of the Zeeman energy at temperature $T/D=5\times 10^{-6}$. By applying a magnetic field to the QD, the zero-bias peak splits up into two peaks of smaller amplitude. The distance between the split peaks is twice the Zeeman energy ($2\Delta\varepsilon_d$). For a reduced value of the Zeeman energy, the split Kondo peaks broaden and begin to overlap. These results are in agreement with previous theoretical studies of QDs attached to metallic leads in the presence of a magnetic field \cite{meir1993,dong2001}.

\begin{figure}
\includegraphics[width=8.2cm,keepaspectratio]{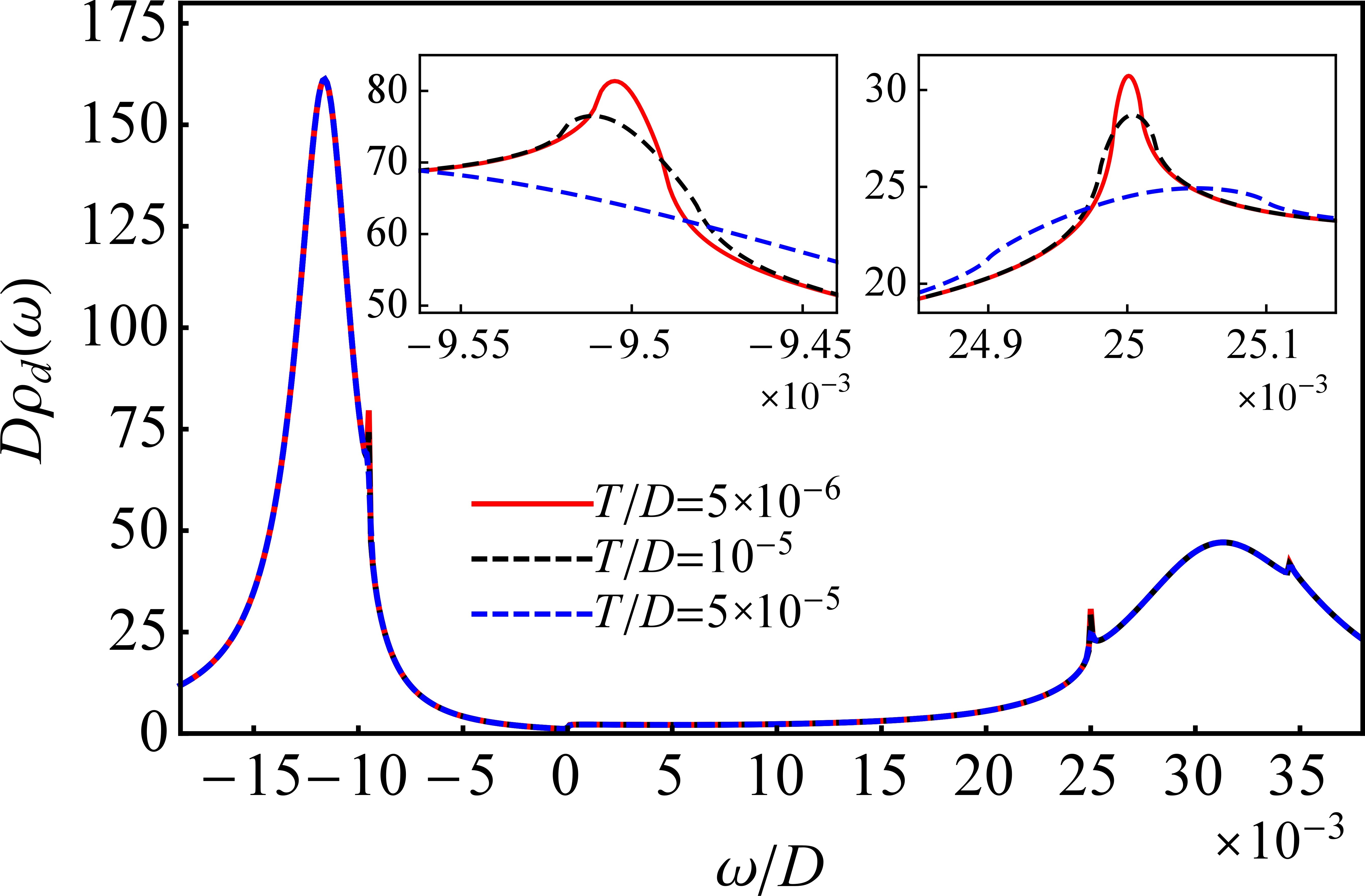}
\caption{The nonequilibrium DOS in the QD for $U/D = 0.069$ with $\mu_L/D=25\times 10^{-3}$ and $\mu_R/D=-9.5\times 10^{-3}$ at three different temperatures and zero magnetic field. Here, $\eta=0.02$ and $\varepsilon_d/D=-0.022$.}
\label{fig:9}
\end{figure}

The numerical results for the nonequilibrium DOS in the QD calculated for finite $U$ at three different temperatures in the absence of a magnetic field are shown in Figure \ref{fig:9}. At low temperature, similar to the case of $U \to \infty$ (see Figure \ref{fig:2} and Figure \ref{fig:3}), the Kondo resonances appear at the chemical potentials, and their amplitudes decrease with increasing temperature and disappear at high temperature. Besides the temperature-dependent narrow and sharp Kondo peaks, the DOS consists of two broadened peaks of large amplitude located at $\omega \approx \textmd{Re}N_1(\omega)$ and $\omega \approx \textmd{Re}N_2(\omega)$, where $\omega - N_1(\omega)$ and $\omega - N_2(\omega)$ are the denominators of the first and the second part of the Green's function of the QD given by Equation \eqref{eq:5}, in absence of magnetic fields. The broadened peaks determine the differential conductance within the Coulomb blockade regime. The left broadened peak corresponds to a resonant transmission through the QD at $\omega \approx \tilde{\varepsilon}_d$ and the right one to $\omega \approx \tilde{\varepsilon}_d + \tilde U$, where the renormalized QD energy level $\tilde{\varepsilon}_d$ and the renormalized Coulomb interaction $\tilde U$ have to be calculated self-consistently using the method presented by Van Roermund et al. in \cite{roermund2010}.

\begin{figure}
\includegraphics[width=8.2cm,keepaspectratio]{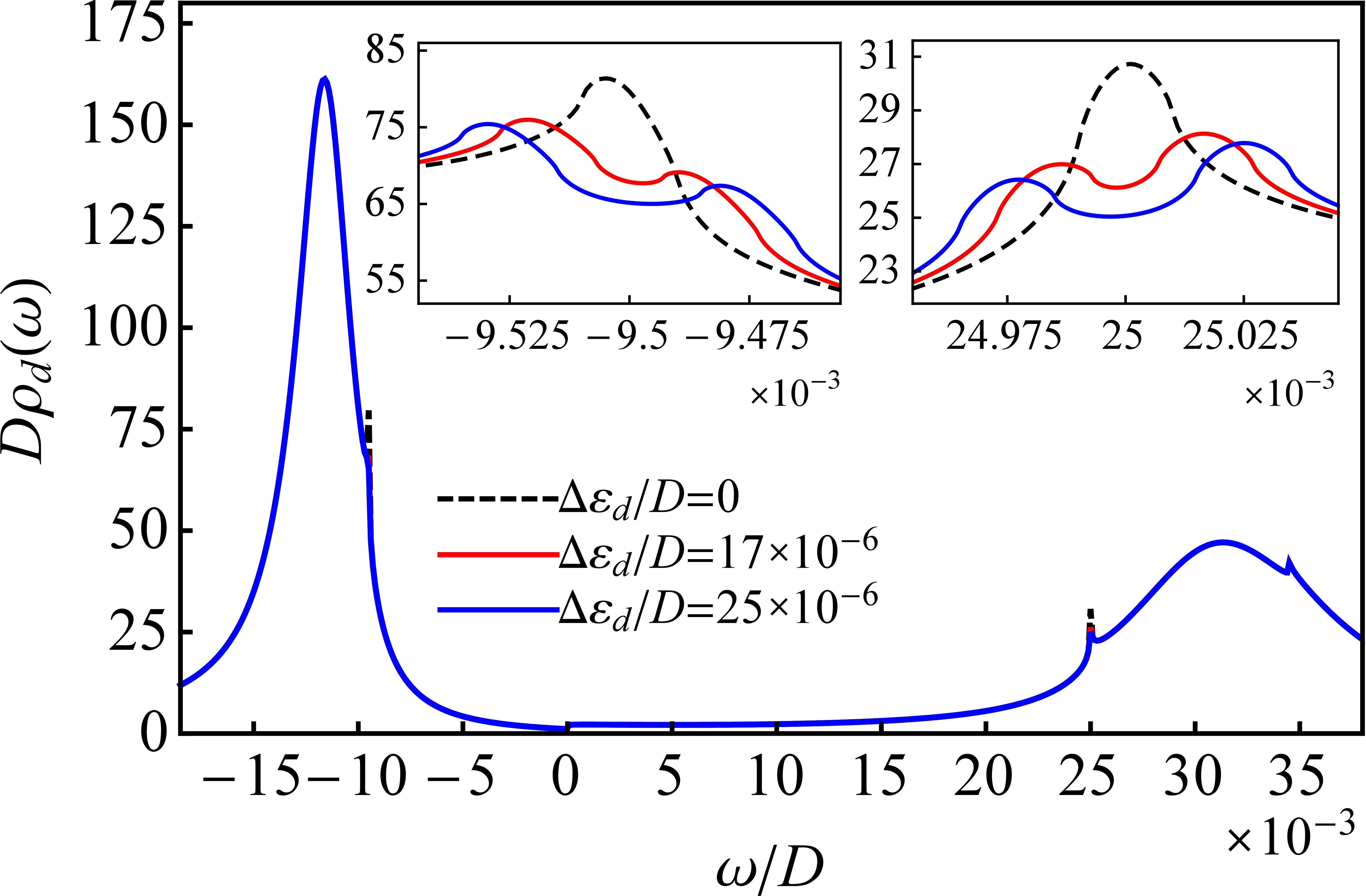}
\caption{The nonequilibrium DOS in the QD for different values of the Zeeman energy and finite $U$ at temperature $T/D=5\times 10^{-6}$. The remaining parameters are the same as those in Figure \ref{fig:9}.}
\label{fig:10}
\end{figure}

Figure \ref{fig:10} shows the self-consistent calculations of the total DOS for finite $U$ at temperature $T/D=5 \times 10^{-6}$ in the presence of a magnetic field. In agreement with the results for $U \to \infty$ (see Figure \ref{fig:4}), the Kondo resonances split up when a magnetic field is applied to the QD. The split peaks have smaller amplitudes and are shifted by the Zeeman energy from the chemical potentials, to the right for spin-up electrons and to the left for spin-down electrons. The damped peaks characterize the transport in the Coulomb blockade regime in the presence of a magnetic field and are located at $\omega \approx \textmd{Re}N_{1\sigma}(\omega)$ and $\omega \approx \textmd{Re}N_{2\sigma}(\omega)$ for electrons with spin $\sigma$. The broadened peaks shift by $\Delta\varepsilon_d/2$ from the locations determined by $\textmd{Re}N_{i}(\omega)$ in absence of magnetic fields, to the right for spin-up electrons and to the left for spin-down electrons. In the same way, the spin-dependent renormalized QD energy level $\tilde{\varepsilon}_{d\sigma}$ and the renormalized Coulomb interaction energy $\tilde U$ can be estimated.

\begin{figure}
\includegraphics[width=8.2cm,keepaspectratio]{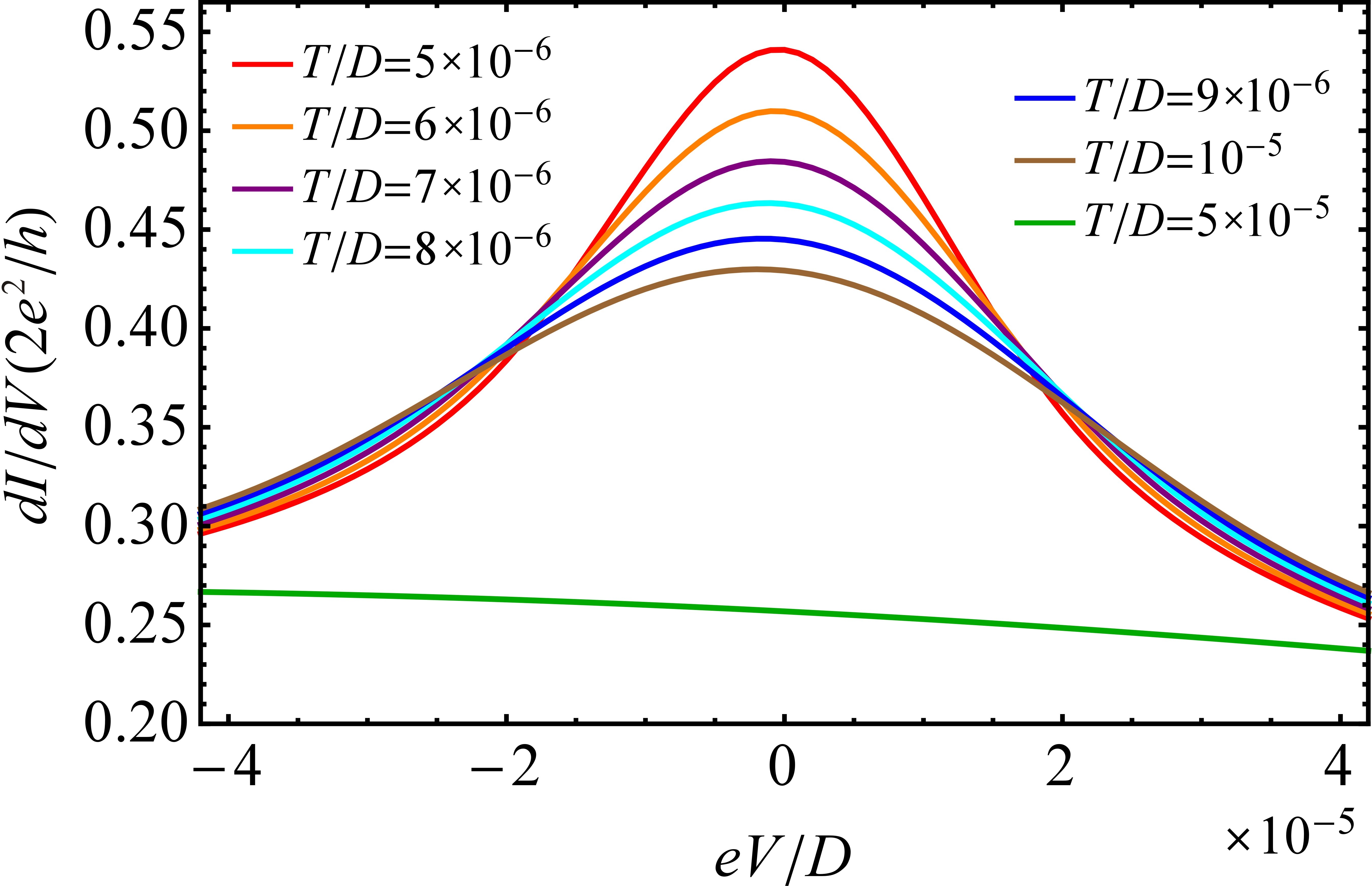}
\caption{Differential conductance $dI/dV$ as a function of the bias voltage $eV$ with $\mu_R/D=-9.5 \times 10^{-3}$ for different temperatures and finite $U$ at zero magnetic field. The remaining parameters are the same as those in Figure \ref{fig:9}.}
\label{fig:11}
\end{figure}

In Figure \ref{fig:11}, we present the calculated total differential conductance as a function of the bias voltage for finite Coulomb interaction and at different temperatures and zero magnetic field. In analogy to the case of $U \to \infty$ (see Figure \ref{fig:5}), the zero-bias peak appears in the plot of the differential conductance when the bias voltage equals the Zeeman energy ($eV=\Delta\varepsilon_d=0$). Its shape strongly depends on the temperature. At high temperature, the zero-bias peak is not observed. With decreasing temperature the Kondo peak appears and becomes increasingly sharper due to the narrow Kondo resonance in the DOS. Therefore, close to $eV=0$, the Kondo peak strongly depends on $eV$. It significantly decreases when $eV$ increases. The variation of the differential conductance for small values of $|eV|$ can be approximated by a second-degree polynomial function as has already been proposed in the $U \to \infty$ case. The $\mu_R$ dependence of the zero-bias peak shows the same behavior as observed for $U \to \infty$.

\section{Conclusion}
\label{sec:IV}
We studied the Kondo effect of a QD connected to pure monolayer graphene electrodes arranged in an armchair configuration with respect to the QD. To the best of our knowledge, this is the first study exploring the influence of a magnetic field on the transport properties of such a system. The system is described by the pseudogap Anderson model under external magnetic fields. We derived an analytical formula for the Green's function of the QD for finite on-site Coulomb interaction using standard decoupling schemes. An analytical formula for the Kondo temperature is also derived for electron and hole doping of the graphene. The Kondo temperature vanishes at the Dirac point close to the particle–hole symmetry point. In the case of particle–hole asymmetry, the Kondo temperature has a nonzero value even at the Dirac point. This behavior is in agreement with other reports in the literature. The DOS of the QD and the differential conductance through the QD were calculated self-consistently for finite temperature. The finite on-site Coulomb interaction has a similar effect on the transport properties as was observed for QDs connected to metallic electrodes. The DOS of the QD is vanishing for zero energy values since the DOS of graphene is zero at the Dirac points. Consequently, the zero-bias peak is not observed when the chemical potential matches the Dirac points of graphene. An analytical method to calculate the integrals appearing in the self-energies is developed, which can be applied for related graphene-based systems.

To the best of our knowledge, there are no experimental studies of the presented system. Therefore, we suggest an experimental realization of the system, as shown in Figure \ref{fig:1}, which will verify our theoretical model. We hope that the results obtained will contribute to the development of new graphene-based nanoelectronic devices. 

\begin{acknowledgements}
We would like to thank Dr. Doru Sticleţ, Dr. Liviu P. Zârbo, Zoltán Kovács-Krausz and Pál-Attila Máthé for valuable discussions. L. M. acknowledges financial support from the Eötvös Loránd University and Ministry of Human Capacities of Hungary through a scholarship for Ph.D. students with the contract number ELTE/5455/-35/2018 and the Romanian Government for a scholarship during the academic year 2017–2018 with the contract number 18500/13013/05.10.2016 and a grant of the Romanian National Authority for Scientific Research and Innovation, CNCS-UEFISCDI, with project number PN-III-P1-1.2-PCCDI-2017-0338.
\end{acknowledgements}

\bibliography{References}
\end{document}